\documentclass[12pt,preprint]{aastex}

\def\p    {\phantom{0}}

\def\Prob {\ifmmode {{\rm Prob}}\else{Prob}\fi}
\def\Prior{\ifmmode {{\rm Prior}}\else{Prior}\fi}

\def\kms  {km~s$^{-1}$}

\def\masd {mas~d$^{-1}$}
\def\masy {mas~y$^{-1}$}

\def\uas  {\ifmmode {\mu{\rm as}}\else{$\mu$as}\fi}
\def\deg  {\ifmmode {^\circ}\else {$^\circ$}\fi}
\def\porm {\ifmmode {\pm}\else {$\pm$}\fi}
\def\chisqpdf {\ifmmode {\chi^2_{\rm pdf}}\else {$\chi^2_{\rm pdf}$}\fi}
\def\chisq    {\ifmmode {\chi^2}\else {$\chi^2$}\fi}
\def\Msun {M$_\odot$}
\def\Mbh  {\ifmmode{M}\else{$M$}\fi}

\def\HII  {H~{\small II}}

\def\etal {et al.~}
\def\eg   {e.g.,~}
\def\ie   {i.e.,~}

\def\d    {\ifmmode {{\rlap{.}}^\circ}\else {${\rlap{.}}^\circ$}\fi}
\def\s    {\ifmmode {{\rlap{.}}^s}\else {${\rlap{.}}^s$}\fi}
\def\as   {\ifmmode {{\rlap{.}}^{''}}\else {${\rlap{.}}^{''}$}\fi}

\newbox\grsign \setbox\grsign=\hbox{$>$} \newdimen\grdimen \grdimen=\ht\grsign
\newbox\laxbox \newbox\gaxbox
\setbox\gaxbox=\hbox{\raise.5ex\hbox{$>$}\llap
     {\lower.5ex\hbox{$\sim$}}}\ht1=\grdimen\dp1=0pt
\setbox\laxbox=\hbox{\raise.5ex\hbox{$<$}\llap
     {\lower.5ex\hbox{$\sim$}}}\ht2=\grdimen\dp2=0pt
\def\gax{\mathrel{\copy\gaxbox}}
\def\lax{\mathrel{\copy\laxbox}}

\def\pa    {\ifmmode {\psi} \else {$\psi$}\fi}
\def\rPpm  {\ifmmode {r_{\Ro,\To}} \else {$r_{Ro,\To}$}\fi}

\def\vlsr  {\ifmmode {v_{\rm LSR}}\else {$v_{\rm LSR}$}\fi}
\def\vlsrr {\ifmmode {v^r_{\rm LSR}}\else {$v^r_{\rm LSR}$}\fi}
\def\vhelio{\ifmmode {v_{Helio}}\else {$v_{Helio}$}\fi}

\def\ura   {\ifmmode {\mu_\alpha}\else {$\mu_\alpha$}\fi}
\def\udec  {\ifmmode {\mu_\delta}\else {$\mu_\delta$}\fi}

\def\ul    {\ifmmode {\mu_l}\else {$\mu_l$}\fi}
\def\ub    {\ifmmode {\mu_b}\else {$\mu_b$}\fi}

\def\uml   {\ifmmode {v_{gr}}\else {$v_{gr}$}\fi}
\def\umb   {\ifmmode {v_b}\else {$v_b$}\fi}
\def\vsrad {\ifmmode {v_{rad}}\else {$v_{rad}$}\fi}

\def\upl   {\ifmmode {v^p_{gr}}\else {$v^p_{gr}$}\fi}
\def\upb   {\ifmmode {v^p_b}\else {$v^p_b$}\fi}
\def\vprad {\ifmmode {v^p_{rad}}\else {$v^p_{rad}$}\fi}

\def\Vo    {\ifmmode {V^{Std}_\odot}\else {$V^{Std}_\odot$}\fi}
\def\Uo    {\ifmmode {U^{Std}_\odot}\else {$U^{Std}_\odot$}\fi}
\def\Wo    {\ifmmode {W^{Std}_\odot}\else {$W^{Std}_\odot$}\fi}
\def\VH    {\ifmmode {V^H_\odot}\else {$V^H_\odot$}\fi}
\def\UH    {\ifmmode {U^H_\odot}\else {$U^H_\odot$}\fi}
\def\WH    {\ifmmode {W^H_\odot}\else {$W^H_\odot$}\fi}
\def\V     {\ifmmode {V_\odot}\else {$V_\odot$}\fi}
\def\U     {\ifmmode {U_\odot}\else {$U_\odot$}\fi}
\def\W     {\ifmmode {W_\odot}\else {$W_\odot$}\fi}

\def\Vs    {\ifmmode {V_s}\else {$V_s$}\fi}
\def\Us    {\ifmmode {U_s}\else {$U_s$}\fi}
\def\Ws    {\ifmmode {W_s}\else {$W_s$}\fi}

\def\Vsbar {\ifmmode {\overline{V_s}}\else {$\overline{V_s}$}\fi}
\def\Usbar {\ifmmode {\overline{U_s}}\else {$\overline{U_s}$}\fi}
\def\Wsbar {\ifmmode {\overline{W_s}}\else {$\overline{W_s}$}\fi}

\def\aone  {\ifmmode {a_1}\else {$a_1$}\fi}
\def\atwo  {\ifmmode {a_2}\else {$a_2$}\fi}
\def\athr  {\ifmmode {a_3}\else {$a_3$}\fi}

\def\pars  {\ifmmode{\pi_s}\else{$\pi_s$}\fi}

\def\To    {\ifmmode{\Theta_0}\else{$\Theta_0$}\fi}
\def\Ro    {\ifmmode{R_0}\else{$R_0$}\fi}
\def\Ho    {\ifmmode{H_0}\else{$H_0$}\fi}

\def\GRS    {GRS~1915+105}
\def\QSON   {J1913+1220}
\def\QSOS   {J1913+0932}

\def\Mref   {G045.07+0.13}
\def\Mcheck {G045.45+0.05}
\def\IRASS  {IRAS~19132+1035}
\def\IRASN  {IRAS~19124+1106}

\shorttitle{The Distance to \GRS} 
\shortauthors{Reid \etal}

\begin{document}

\title{A Parallax Distance to the Microquasar \GRS\ and a Revised
       Estimate of its Black Hole Mass}

\author{M. J. Reid\altaffilmark{1},
        J. E. McClintock\altaffilmark{1}, 
        J. F. Steiner\altaffilmark{1},
        D. Steeghs\altaffilmark{3},
        R. A. Remillard\altaffilmark{2},
        V. Dhawan\altaffilmark{4},
        R. Narayan\altaffilmark{1}
       }

\altaffiltext{1}{Harvard-Smithsonian Center for
   Astrophysics, 60 Garden Street, Cambridge, MA 02138, USA}
\altaffiltext{2}{MIT Kavli Institute for Astrophysics and Space Research, MIT, 
   70 Vassar Street, Cambridge, MA 02139, USA}
\altaffiltext{3}{Department of Physics, The University of Warwick, Coventry CV4 7AL, UK}
\altaffiltext{4}{National Radio Astronomy Observatory, PO Box 0, Socorro, NM, 87801, USA}

\begin{abstract}
Using the Very Long Baseline Array, we have measured a trigonometric parallax for 
the microquasar \GRS, which contains a black hole and a K-giant companion.  
This yields a direct distance estimate of $8.6^{+2.0}_{-1.6}$ kpc and a revised 
estimate for the mass of the black hole of $12.4^{+2.0}_{-1.8}$ \Msun.  
\GRS\ is at about the same distance as some \HII\ regions 
and water masers associated with high-mass star formation in the Sagittarius spiral 
arm of the Galaxy.  The absolute proper motion of \GRS\ is $-3.19\pm0.03$ \masy\ 
and $-6.24\pm0.05$ \masy\ toward the east and north, respectively, 
which corresponds to a modest peculiar speed of 
$22\pm24$ \kms\ at the parallax distance, suggesting that the binary did not 
receive a large velocity kick when the black hole formed.  
On one observational epoch, \GRS\ displayed superluminal motion 
along the direction of its approaching jet.  Considering previous observations of 
jet motions, the jet in \GRS\ can be modeled with a jet inclination to the line of 
sight of $60^{\rm \circ}\pm5^{\rm \circ}$ and a variable flow speed between $0.65c$ 
and $0.81c$, which possibly indicates deceleration of the jet at distances from the 
black hole $\gtrsim2000$~AU.  Finally, using our measurements of distance and estimates 
of black hole mass and inclination, we provisionally confirm our earlier result that 
the black hole is spinning very rapidly.
\end{abstract}


\section{Introduction} \label{sect:intro}

The Galactic microquasar, \GRS, is a low-mass X-ray binary containing a black hole 
and a K~{\footnotesize III} companion \citep{Greiner:01} with a 34-day orbital period.
The companion overflows its Roche lobe and the system exhibits episodic superluminal 
radio jets \citep{Mirabel:94}.  The estimated nature of the companion 
(\eg\ mass, luminosity, size) and the mass and spin of the black hole depend on 
the distance to the system, which is highly uncertain.
A luminosity distance based on the spectral identification of the companion
is hindered by the large extinction, $A_v \approx 25-30$ mag \citep{Chapuis:04}, 
and the uncertainties in stellar radius and surface brightness of the tidally-distorted
secondary.  Alternatively, a kinematic distance estimate would rely on the 
line-of-sight velocity of the system, assuming that the binary executes
a circular orbit in the Galaxy.  This is a weak assumption for a system
that may be a Gyr old and whose velocity could have been significantly affected by 
mass ejection during the formation of the black hole.

\citet{Fender:99} determined an upper limit on distance of 12.5 kpc (95\% confidence, based 
on the quoted limit of $11.2\pm0.8$ kpc) by observing approaching and receding jet motions 
and requiring that the bulk flow speed in the jet not exceed the speed of light.  Absorption of 
the \GRS\ continuum emission by interstellar hydrogen at 21-cm wavelength indicates significant 
opacity over $0 < \vlsr < 75$ \kms \citep{Mirabel:94}.  This is consistent with \GRS\ 
being more distant than the Galactic tangent point of 5.9 kpc at its Galactic longitude 
of $45\d37$, where \vlsr\ peaks at 70 \kms\ for $\Ro=8.34$ kpc and $\To=240$ \kms\  
\citep{Reid:14}.   Thus, these two limits only constrain the distance of \GRS\ to between 
5.9 and 12.5 kpc.  In order to obtain a solid distance measurement for \GRS, 
we undertook a series of observations with the National Radio Astronomy 
Observatory's\footnote{The National Radio Astronomy Observatory is a facility of the 
National Science Foundation operated under cooperative agreement by Associated 
Universities, Inc.} 
Very Long Baseline Array (VLBA) to measure a trigonometric parallax.

Unlike the black hole binaries V404~Cyg \citep{Miller-Jones:09} and Cyg~X-1 \citep{Reid:11}, 
for which measurements of trigonometric parallax were relatively straightforward,
\GRS\ is a difficult target for several reasons.  Firstly, compact radio emission
from the base of the jets, associated with the X-ray hard state, occurs only
about 25\% of the time.  Secondly, when in the hard state, the source 
is typically weak, $<10$ mJy at cm-wavelengths, and often falls below astrometrically
useful flux densities of $\lax1$ mJy.  Owing to these difficulties, a long series of
observations were required, as documented in Section \ref{sect:observations}.
Thirdly, on some occasions, bright knots of radio emission propagate down the jets, 
leading to outlying data points when fitting for parallax and proper motion.  
Such behavior was clearly observed on one epoch, as described in Section \ref{sect:jetmotion}.
These data are interesting for studies of the jet, but cannot be used for parallax
fitting.  Finally, the apparent core position exhibits some ``jitter'' along the 
jet axis, probably owing to the production of low-level jet emission within our
angular resolution of $\sim1$ mas.  This led us to fit for the parallax shifts
with the data rotated parallel and perpendicular to the jet axis, instead of the 
standard right ascension and declination directions, in order to apply optimal weighting.

The parallax distance measurement of \GRS\ is described in Section \ref{sect:parallax} and was 
accomplished in two steps: 1) measuring a relative parallax distance with respect to a Galactic 
water maser projected within 0\d5 of \GRS\ on the sky, and 2) measuring an absolute 
parallax distance for the water maser with respect to background quasars.   
In Section \ref{sect:discussion}, we discuss the astrophysical implications of our 
astrometric results, while providing revised estimates of the jet parameters and 
the black hole's mass and spin.

\section{Observations \& Results} \label{sect:observations}

In total, we observed \GRS\ on 43 epochs at 22.235 GHz under VLBA programs BM257 and BR145.
The source was detected ($>1$ mJy/beam) on 18 epochs.  Seventeen detections are reported in 
Table~\ref{table:observations} and analyzed for parallax and proper motion in 
Section \ref{sect:parallax}.  
The remaining observation displayed a rapidly moving component and is discussed in
Section \ref{sect:jetmotion} 
  
\begin{figure}[ht]
\epsscale{0.7} 
\plotone{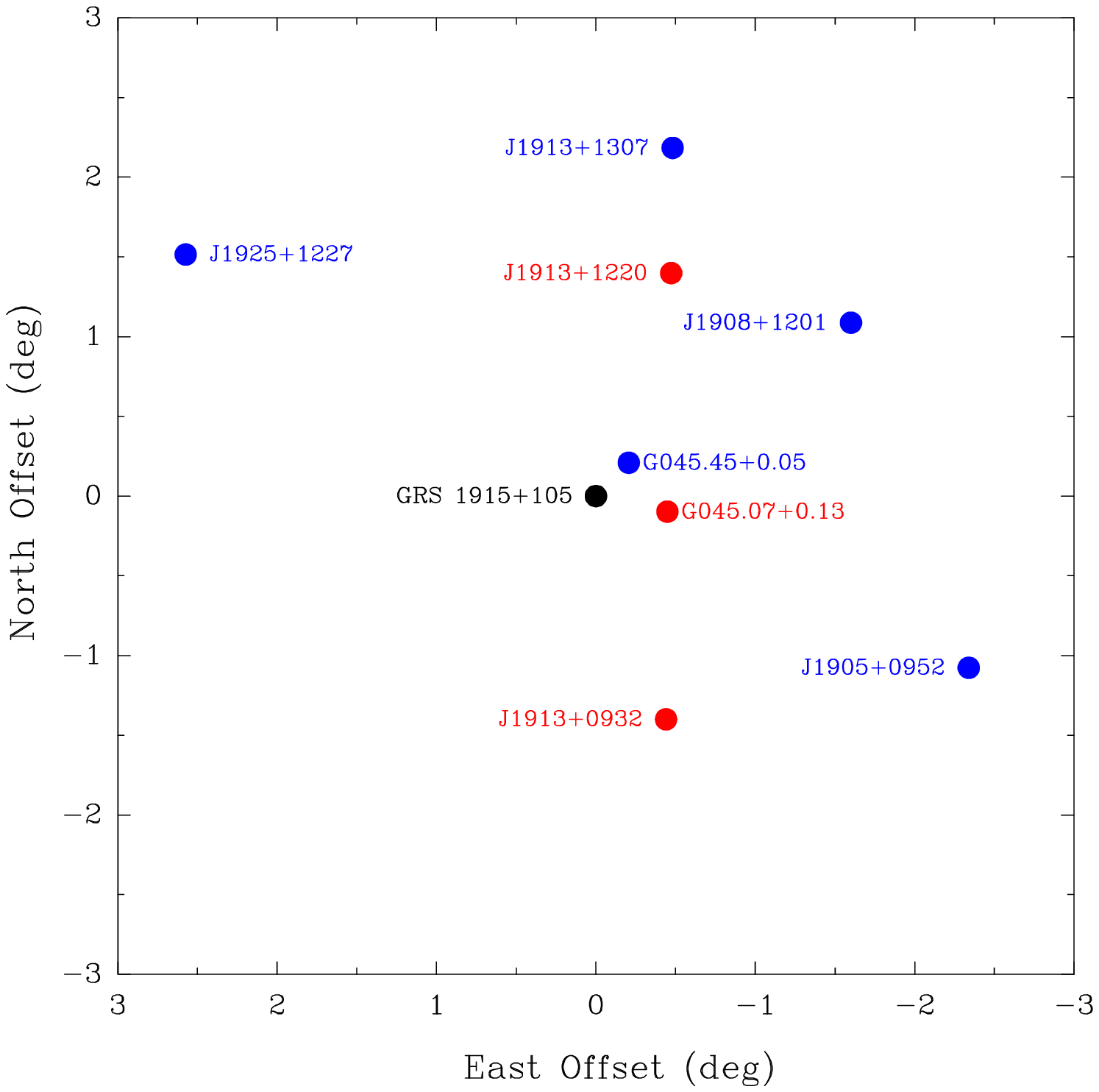}
\caption{\footnotesize
~Relative locations of sources used for parallax observations.  The Galactic water maser
\Mref\ was used as the interferometer phase-reference for \GRS, \QSOS\ and \QSON.
In this paper we determine the relative parallax of \GRS\ (black) with respect to \Mref,
(red).   The absolute parallax of \Mref\ relative to \QSOS\ and \QSON\ (red) 
and the absolute parallax of a nearby water maser, \Mcheck, relative to four quasars (blue)
are documented by \citet{Wu:14} and discussed in Section \ref{sect:parallax}.
         }
\label{fig:locations}
\end{figure}

A Galactic water maser associated with a star forming region in the Sagittarius spiral
arm of the Galaxy was used as the interferometer phase-reference \citep{Wu:14}.
A typical observation sequence was \Mref, \GRS, \Mref, \QSOS, \Mref, \QSON,
switching sources every 20-30 seconds and repeating this sequence for a period of 
12 minutes.  The relative locations of these sources, as well as another water maser, \Mcheck,
and the four quasars used to measure its parallax are shown in Fig. \ref{fig:locations}.

We recorded data at a rate of 512 Mbits per second with 
four intermediate frequency bands of 16 MHz in right and left circular 
polarizations, with the \Mref\ maser centered in the second band.
In order to improve astrometric accuracy, we inserted 3 or 4 half-hour long 
``geodetic blocks'' designed to measure clock drifts and residual tropospheric delays 
not removed in the correlation process.  Corrections for these residual delays 
were done in post-correlation calibration.  Details of the calibration
of astrometric VLBA data are given in \citet{Reid:09}.

\begin{deluxetable}{crrc}
\tablecolumns{4} \tablewidth{0pc} 
\tablecaption{\GRS\ Parallax Data}
\tablehead {
 \colhead{Epoch} &\colhead{East Offset} &\colhead{North Offset} &\colhead{Brightness} \\
 \colhead{(yr)}  &\colhead{(mas)}       &\colhead{(mas)}        &\colhead{(mJy/beam)}  
           }
\startdata
  2008.333       &$0.586\pm0.003$       &$0.288\pm0.006$        &35.2                  \\
  2008.335       &$0.579\pm0.003$       &$0.298\pm0.006$        &33.4                  \\
  2009.220       &$0.443\pm0.010$       &$0.700\pm0.018$        &\p4.6                 \\
  2009.229       &$0.407\pm0.014$       &$0.894\pm0.028$        &\p3.1                 \\
  2009.733       &$0.276\pm0.027$       &$0.886\pm0.037$        &\p3.3                 \\
  2009.835       &$0.118\pm0.016$       &$0.962\pm0.030$        &\p2.4                 \\
  2009.843      &$-0.052\pm0.045$       &$1.085\pm0.073$        &\p1.1                 \\
  2010.197       &$0.001\pm0.009$       &$1.022\pm0.014$        &12.1                  \\
  2010.329      &$-0.113\pm0.007$       &$1.077\pm0.013$        &10.1                  \\
  2010.698       &$0.992\pm0.020$      &$-0.314\pm0.038$        &\p5.0                 \\
  2010.810      &$-0.097\pm0.020$       &$1.053\pm0.035$        &\p3.2                 \\
  2010.873       &$0.654\pm0.076$       &$0.175\pm0.074$        &\p3.1                 \\
  2011.862      &$-0.357\pm0.009$       &$1.373\pm0.016$        &20.6                  \\
  2011.871      &$-0.407\pm0.015$       &$1.350\pm0.025$        &12.0                  \\
  2013.019      &$-0.487\pm0.021$       &$1.695\pm0.050$        &\p3.7                 \\
  2013.153      &$-0.392\pm0.006$       &$1.316\pm0.009$        &14.7                  \\
  2013.255      &$-0.444\pm0.022$       &$1.450\pm0.055$        &\p3.7                 \\
\enddata
\tablecomments{\footnotesize
~Epochs of observation when \GRS\ was detectable ($>1$ mJy), along with measured position 
offsets of \GRS\ with respect to a \Mref\ $\vlsr=60$ \kms\ maser spot (with J2000 positions
for \GRS\ of ($19^h15^m11\s5473,+10^d56'44\as704$) and \Mref\ of ($19^h13^m22\s0427,+10^d50'53\as336$)
used in the VLBA correlator)  and the observed peak brightness of \GRS.  
Position errors are formal $1\sigma$ fitting uncertainties.  
Position offsets from epochs 2010.698 and 2010.873 were outliers and not used in the 
parallax fitting (see Section \ref{sect:parallax}).  
              }
\label{table:observations}
\end{deluxetable}

We used the maser emission at $\vlsr=60$ \kms\ as the interferometer phase reference.
Power in this spectral channel of width 0.42 \kms\ came from two maser spots separated by
about 6 mas.  Over the course of the observations, one spot weakened and the other strengthened.
This shifted the phase reference position over time.  We imaged both spots and were able to
measure the position of the originally brightest spot at all epochs.  This served as the
position reference for the parallax data.
Calibrated VLBA data were imaged with the Astronomical Image Processing System (AIPS)
task {\footnotesize IMAGR}.  The brightness and position offsets of \GRS\ (relative
to the maser), measured by fitting elliptical Gaussian brightness distributions with 
the task {\footnotesize JMFIT}, are listed in Table \ref{table:observations}.  

\subsection{Jet Motions in \GRS} \label{sect:jetmotion}

\begin{figure}[ht]
\epsscale{0.7} 
\plotone{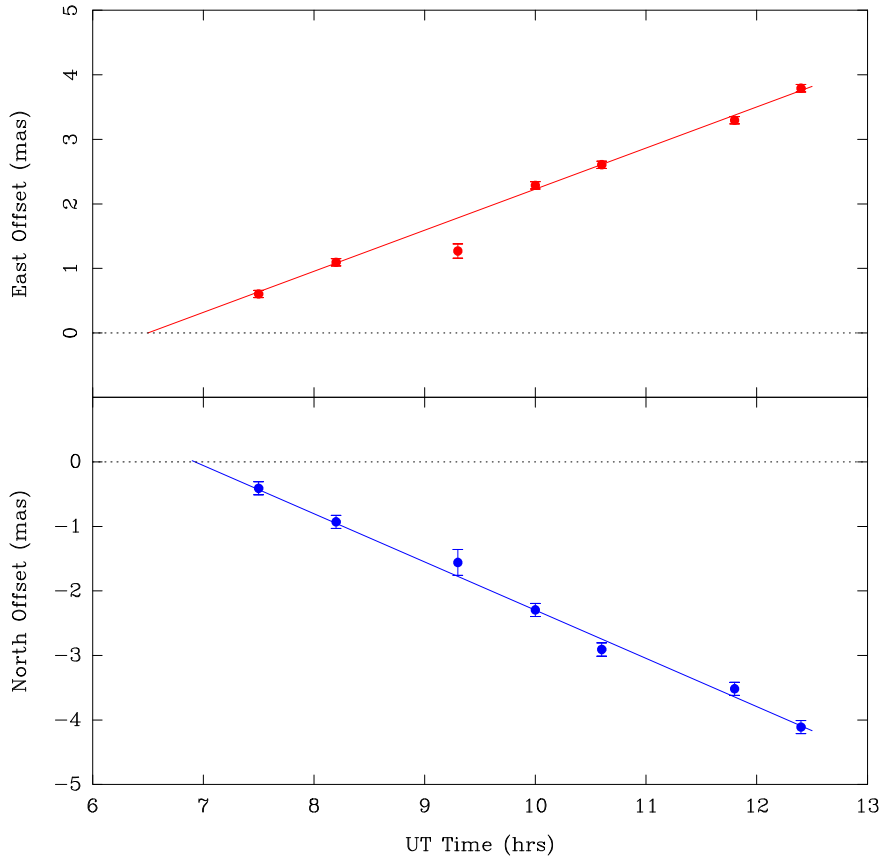}
\caption{\footnotesize
~Eastward (red) and Northward (blue) offsets of the peak emission of \GRS\
as a function of time on 2013 May 24.  The Galactic water maser \Mref\ was used as 
the interferometer phase-reference and zero offsets (dotted lines) are the expected 
position of the \GRS\ core based on the parallax and proper motion fitting 
(see Section \ref{sect:parallax}). The apparent outlier at UT=9.3 hr was not used 
when fitting for the motion components (solid lines).
         }
\label{fig:jetmotion}
\end{figure}

On 2013 May 24, \GRS\ was strongly variable and an image 
showed very complex structure extended over about 5 mas in the NW--SE direction.
This orientation is similar to that of the jet seen on  $\sim1''$ scales
with interferometric imaging by \citet{Mirabel:94} using the VLA,
on 250 mas scales by \citet{Fender:99} using the MERLIN array,
and on 5--50 mas scales by \citet{Dhawan:00} using the VLBA,
when \GRS\ exhibited superluminal expansion.  Since jet motion is expected to be of 
order a synthesized beam size per hour, we re-imaged the data from individual 
20-min spans.  
We used the 8 continental antennas of the VLBA (excluding the very long baselines
to the Mauna Kea antenna and the St. Croix antenna, which generally suffers
from high water vapor opacity) to produce images.  Owing to the limited 
interferometer (u,v)-coverage afforded by these ``snap shot'' images, we could only reliably 
measure the position of the peak emission, which we restricted to be within $\pm10$ mas of the 
expected core position based on the parallax and proper motion fitting results 
(see Section \ref{sect:parallax}).
The position offsets, relative to the expected position of the core of \GRS,
are shown in Fig. \ref{fig:jetmotion}.

Both the eastward and northward offsets of the peak emission versus time display
linear motion, with one apparent outlier at UT=9.3 hr.  Removing the outlier and 
fitting a constant motion to these data yields $23.6\pm0.5$ \masd\ toward 
$130.5\deg\pm1.1\deg$ east of north.  
Since, these positions were registered relative to the expected
position of the core (black hole) of \GRS, the zero crossings should indicate the
time of the expulsion of the radiating plasma.  This occurs near UT 6.5 hr and 6.9 hr
for the easterly and northerly measurements, respectively, suggesting expulsion at 
UT $6.7\pm0.2$ hr.

\subsection{Parallax Distance of \GRS} \label{sect:parallax}

Parallax measurements at centimeter wavelengths with very long baseline interferometry
involve rapid switching between the target source and one or more background
quasars.  This serves two purposes: extending the coherence time of the interferometer 
by removing short term ($\sim$minute) fluctuations of propagation delay through the 
troposphere (usually caused by variable water vapor) and providing a stationary
position reference for the parallax fitting \citep{Reid:13}.  
While the background quasar can serve both purposes, one can use a different 
source to extend coherence times.  For \GRS\ we used an unusually stable Galactic water 
maser, \Mref, which was fortuitously close on the sky, projected 
within $0\d5$.  Switching between these two sources provided excellent phase 
calibration and {\it relative} parallax data. 

Preliminary fitting for a relative parallax and proper motion to the eastward and northward
offsets revealed two significant ($>4\sigma$ in each coordinate) outliers at 
epochs 2010.698 and 2010.873.  For each case, compared to the previous epoch's
position, these points were offset along a position angle of 
$\approx130\deg\pm4\deg$ east of north, suggesting that low-level jet activity was 
affecting the apparent core position.  This was dramatically confirmed on 2013 May 24 
when a superluminal jet component moving along position angle $130.5\deg\pm1.1\deg$ 
dominated the emission (see Section \ref{sect:jetmotion}).

Generally for VLBI measurements, formal position precision does not fully represent 
astrometric accuracy.  Typically, slowly varying ($\sim$hours) uncompensated differences 
in tropospheric propagation delays among antennas and sometimes structural changes in the 
radio sources among epochs lead to degraded position accuracy \citep{Reid:13}.  
In order to account for
these effects, one generally adds ``error floors'' in quadrature with the formal
measurement uncertainties in both coordinates when fitting for parallax.   For the \GRS\ data,
owing to the likelihood of increased ``jitter'' in the apparent position of the core
\footnote{Note that that the expected ``jitter'' in the position of the black hole
owing to its 33.8 day orbit about the center of mass of the binary is expected to be 
very small: $\sim0.004$ mas} 
along the jet axis at some epochs, we rotated the east and north offset data by
$40\deg$ counter-clockwise on the sky, orienting the approaching jet toward
the negative y-axis.  The rotation angle represents an average jet direction for our 
observations, based on the measured motion in Section \ref{sect:jetmotion}, and those 
of \citet{Mirabel:94} and \citet{Fender:99}.  

Rotating the data allows us to fit for parallax and proper motion with data perpendicular 
and parallel to the jet.  We can then add error-floor values in quadrature with the 
formal measurement uncertainties and adjust their values to achieve component 
chi-squares per degree of freedom ($\chi^2_\nu$) near unity in each coordinate.  
In this way, the more jittery data parallel to the jet receives a larger total
uncertainty and is appropriately down-weighted compared to the more stable data 
perpendicular to the jet.  After removing the two outliers discussed above, 
we achieved $\chi^2_\nu=1.0$ for each component of the data with error floors of 
0.09 and 0.19 mas for the position residuals perpendicular and parallel to the jet, 
respectively.  Given the (expected) larger uncertainties in the parallel compared to the
perpendicular data, and the shape of the parallax ellipse which has a smaller amplitude
in the parallel compared to the perpendicular direction, the parallax for \GRS\ is
nearly totally constrained by the perpendicular-to-the-jet data. 

\begin{figure}[ht]
\epsscale{0.85} 
\plotone{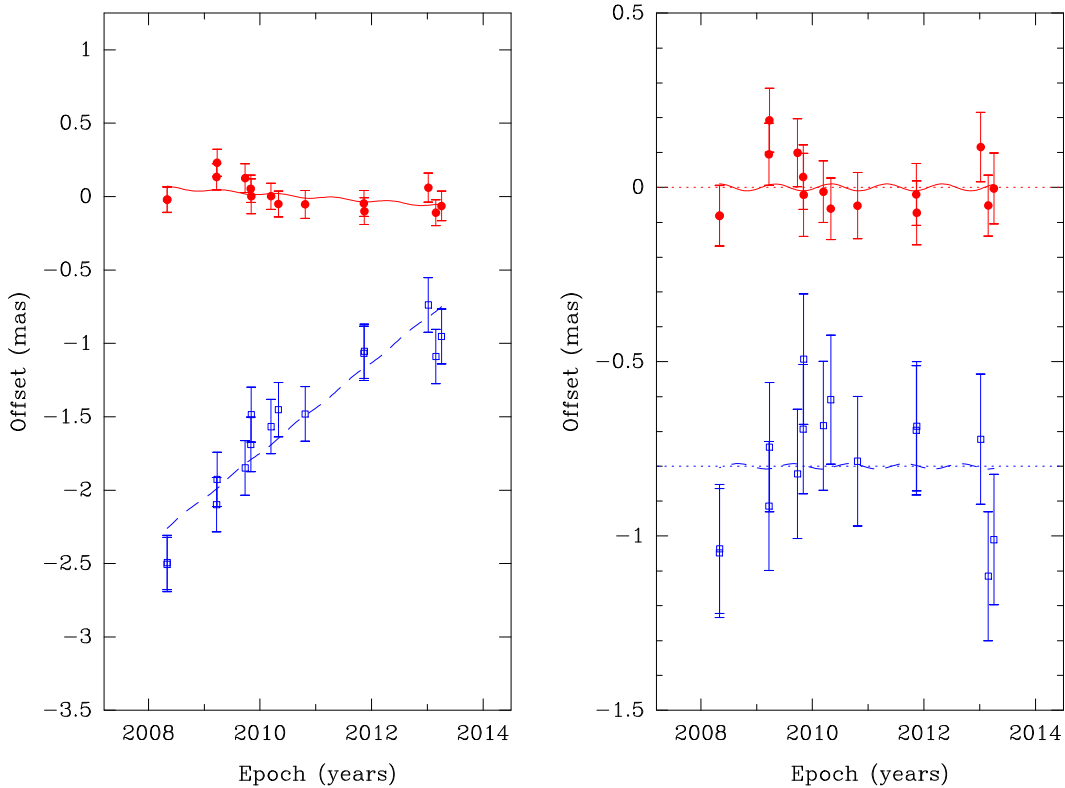}
\caption{\footnotesize
~Relative parallax and proper motion of \GRS\ with respect to the Galactic water 
maser \Mref.
Left panel: Position offsets perpendicular to (filled red circles) and along 
(open blue squares) the approaching jet direction ($140\deg$ east of north)
versus time.  The parallax and motion model is shown with solid and dashed lines.  
The offsets along the jet have been shifted down in the plot for 
clarity.  Right panel: Same as left panel with proper motion removed to display
only the parallax effect. Note that the period (1 y) and phases of the parallax
effect in both coordinate directions are fixed by the Earth's orbit and the
source coordinates; only the amplitude (\ie\ the parallax angle) is a free
parameter.  Since the best fit relative parallax and proper motion components are  
small, \GRS\ and \Mref\ must be at nearly the same distance.
         }
\label{fig:parallax}
\end{figure}

Fitting for the parallax of \GRS\ {\it relative} to the Galactic water maser source \Mref,
with no prior constraints on distance, gives $0.005\pm0.030$ mas, which places both 
sources at nearly the same distance.  Also, the small relative proper motion of 0.3 \masy\ 
is consistent with a similar distance.  Fig. \ref{fig:parallax} displays the  
relative parallax and proper motion for both the data and the model.  
Changing the position angle rotation of the 
data by $\pm5\deg$ changes the relative parallax by about $\mp0.008$ mas.  Thus, 
the relative parallax is not very sensitive to the adopted orientation of the jet axis. 

\begin{figure}[ht]
\epsscale{0.85} 
\plotone{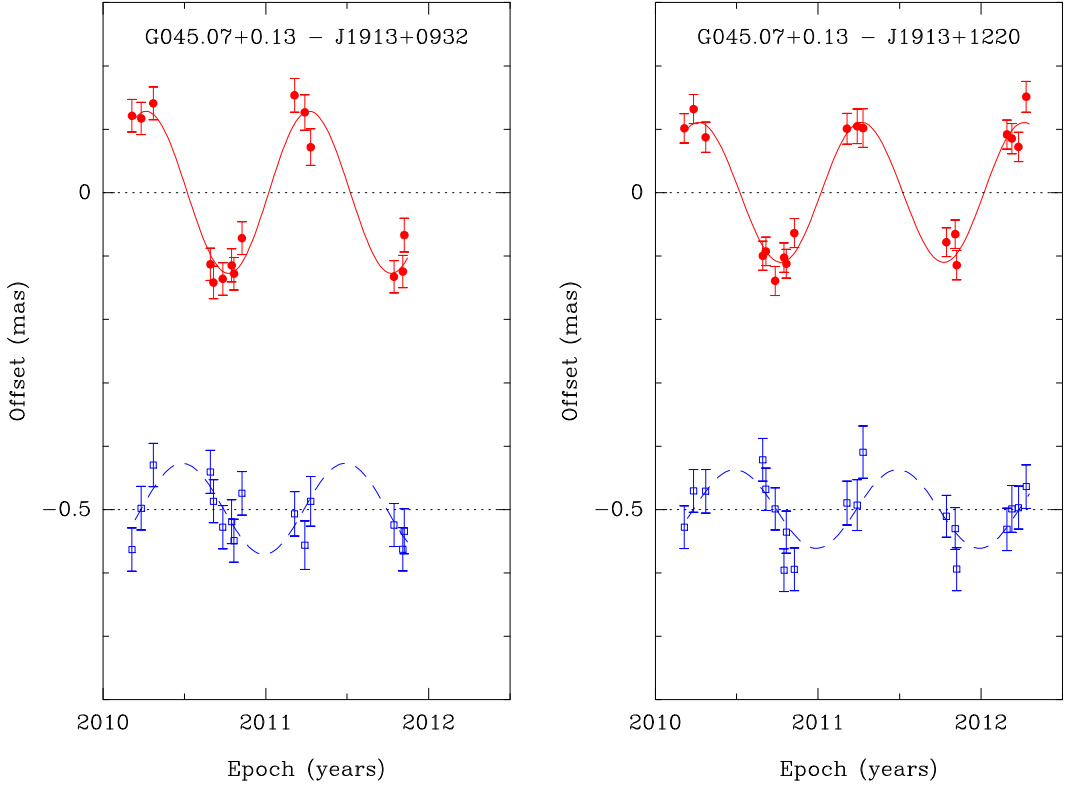}
\caption{\footnotesize
~Absolute parallax of \Mref\ with respect to two background quasars. 
Easterly (filled red circles) and northerly position offsets 
(open blue squares) versus time.  The northerly data have been shifted down
in the plot for clarity.  Proper motions have been removed to display only
the parallax effect.
Left panel: parallax data for \Mref\ vs. \QSOS.
Right panel: parallax data for \Mref\ vs. \QSON.
         }
\label{fig:absolute_parallax}
\end{figure}

In order to obtain an absolute parallax, we also switched between the water maser 
\Mref\ and two quasars, \QSOS\ and \QSON, during the same observations as for \GRS.  
These data were originally presented by \citet{Wu:14}, who reported
a parallax for \Mref\ of $0.129\pm0.007$ mas measured relative to \QSOS.  The parallax
fitting for \Mref\ relative to \QSON\ revealed small systematic deviations from
the model parallax and proper motion.   These deviations could be removed by adding 
position acceleration parameters ($\sim0.1$ mas y$^{-2}$) and gave a parallax estimate of 
$0.111\pm0.007$ mas.  We reproduce the parallax fit results of Wu \etal\ in Fig. 
\ref{fig:absolute_parallax}.  Wu \etal chose to discard this alternative parallax estimate.
Since the two quasars symmetrically straddle \Mref\ on the sky
(see Fig. \ref{fig:locations}), and hence would well cancel most sources of
systematic error, we instead adopt the unweighted average of the two results.  
This yields a parallax for the water maser of $0.120\pm0.009$ mas.

As shown in Fig. \ref{fig:locations}, there is a second Galactic water maser,
\Mcheck, projected within $0\d4$ ($\approx50$ pc) on the sky of \Mref.  
\Mcheck\ has a similar LSR velocity as \Mref, and both masers can be confidently 
associated with the Sagittarius spiral arm of the Milky Way, based on identification
of the masers with a continuous arc of CO emission on Galactic longitude-velocity 
plots \citep{Wu:14}.  Thus, both masers are likely associated with the same 
giant molecular cloud (or adjacent clouds in the same spiral arm) 
and are very likely within about 100 pc of each other.  The parallax 
of \Mcheck\ is $0.119\pm0.017$ mas, measured with respect to the four quasars shown
in Fig. \ref{fig:locations}, which then provides an independent check on the distance 
to \Mref.   As such, we confidently adopt the parallax of \Mref\ of $0.120\pm0.009$ mas, 
corresponding to a distance of $8.33^{+0.68}_{-0.58}$ kpc to this water maser.

For our final estimate of the distance to \GRS, we fit the data with a Bayesian Markov 
chain Monte Carlo (MCMC) approach using the Metropolis-Hastings algorithm to accept or 
reject trials.  We adopted as prior information two constraints on distance ($d$):
1) $d$ greater than the Galactic tangent point distance of 5.9 kpc, as discussed in 
Section \ref{sect:intro}; and
2) $d$ less than 12.0 kpc, where the observed jet motions would require bulk plasma 
flow speeds exceeding that of light.  Rather than use the inclination-dependent
constraint of \citet{Fender:99} of 12.5 kpc (95\% confidence), here we adopt the 
slightly stronger and more direct (inclination independent) limit given by 
$d<c/\sqrt{(\mu_a \mu_r)}$, where $c$ is the speed of light
and $\mu_a$ and $\mu_r$ are the apparent angular speeds of jet components in the
approaching and receding directions.  Thus, we adopt $5.9 < d < 12.0$ kpc with uniform
probability as prior information when fitting the data.   

Traditionally, when fitting relative position as a function of time, one parameterizes 
the model with a parallax parameter (and proper motion terms).  Owing to the
inverse relation between distance and parallax, $\pi$, a flat distance 
prior transforms to a quadratic parallax prior: $\Prior(\pi) = \Prior(d)\times d^2$
(where $\Prior(d)=1$ for $5.9 < d < 12.0$ kpc and zero elsewhere).
Then after fitting, the posteriori probability density function (PDF) 
for parallax should be transformed to
the desired distance PDF, rescaling by $\Prob(d) = \Prob(\pi)\times \pi^2$
(in order to account for the non-linear stretching of $d$ with decreasing $\pi$).
This latter scaling is important only if the parallax uncertainty is a significant
fraction (\eg\ $\gax15$\%) of the parallax.  Alternatively, one can fit the data with 
a model parameterized with distance instead of parallax.  This is more convenient when 
applying prior information on distance and gives the desired distance PDF directly
without rescaling.  Of course, both approaches are equivalent.

\begin{figure}[ht]
\epsscale{0.5} 
\plotone{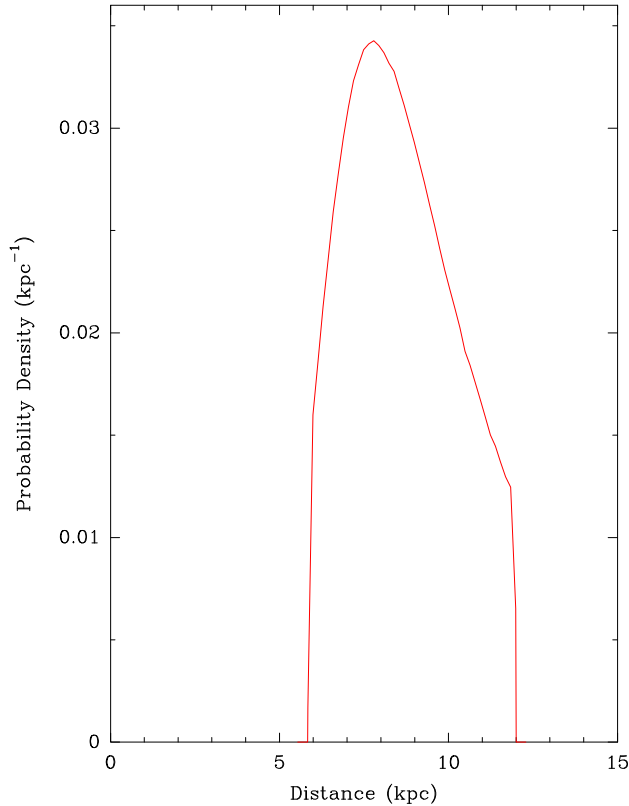}
\caption{\footnotesize
~Probability density function for the parallax distance to \GRS.  The line shows
the probability density obtained by fitting the position vs. time information,
assuming a flat distance prior between the lower (5.9 kpc, from HI absorption 
observed to the Galactic tangent point velocity) and upper (12.0 kpc, based on the 
jet flow speed less than the speed of light) limits for distance, as discussed in 
Section \ref{sect:intro}.
         }
\label{fig:distpdf}
\end{figure}

We adopted the distance parameterization for our final fits and obtain the
posteriori PDF for the distance of \GRS\ shown in Fig.~\ref{fig:distpdf}.
The peak of the PDF is at 7.8 kpc, and the expectation value for distance
is 8.6 kpc, reflecting the asymmetric tail of the PDF toward larger distances.
The minimum-width 68\% confidence interval is 7.0 to 10.6 kpc, which includes the
small contribution from the distance uncertainty of the maser \Mref.   
Adopting the expectation value as the best estimate of distance, we therefore find
\GRS\ at a distance of $8.6^{+2.0}_{-1.6}$ kpc.

\subsection{Proper Motion of \GRS} \label{sect:propermotion}

After removing the effects of the parallax of \GRS, we fit the position offsets 
of \GRS\ with respect to \QSOS\ (without rotating them to align with the jet axis) 
to estimate the proper motion of \GRS\ in the easterly  and northerly directions 
to be $-3.19\pm0.03$ \masy\ and $-6.24\pm0.05$ \masy.
This compares reasonably with the measurement of $-2.86\pm0.07$ \masy\ and 
$-6.20\pm0.09$ \masy\ from 10 years of monitoring with the VLBA by \citet{Dhawan:07}. 

We revisited the systemic velocity of the system whose accuracy, as reported in 
Steeghs et al. (2013), is limited by the published radial velocity values of the 
template stars used.  We retrieved several archival spectra of one of our templates,
HD~176354, which was observed with the high-resolution HARPS spectrograph
at ESO.  Four observations that span four years show a very stable
velocity for this template of $-30.8\pm0.8$ \kms.  This translates
into a refined heliocentric systemic velocity for \GRS\ of 
$\gamma=+12.3 \pm 1.0$~\kms.  Adopting this heliocentric velocity of the \GRS\ binary  
completes the 3-dimensional velocity information.  

Using the Galactic and Solar Motion parameters from \citet{Reid:14} 
($\Ro=8.34$ kpc, $\To=240$ \kms, $\U=10.7, \V=15.6, \W=8.9$ \kms), 
we calculate the peculiar motion of \GRS\ with respect to a circular Galactic orbit.  
We find (\Us,\Vs,\Ws) = ($19\pm3$,$-10\pm24$,$6\pm2$) \kms, where \Us\ is toward
the Galactic center at the location of \GRS, \Vs\ is in the direction of Galactic
rotation, and \Ws\ is toward the north Galactic pole.  This translates to
a non-circular (peculiar) speed of $22\pm24$ \kms\ and a 95\% confidence upper limit
of 61 \kms.   

\section{Discussion }   \label{sect:discussion}

\subsection{Jet Parameters}
The Lorentz factors of ballistic jets from black hole microquasars are generally 
assumed to be in the vicinity $2 \lesssim \Gamma \lesssim 10$ \citep{Fender:04,Fender:06}, 
although there is a dearth of strong constraints.  Specifically, the bulk of 
observational data provide only {\em lower} limits on the Lorentz factor, when 
derived directly from kinematic measurements and flux ratios.  Generally, these lower 
limits fall near $\Gamma_{\rm min} \sim 1.5 - 2$ \citep{Steiner:12a,Steiner:12b,Fender:04}.
As pointed out by \citet{Fender:03}, 
proper motion data are frequently unable to place significant upper bounds on $\Gamma$.  
This is due to the strong, nonlinear degeneracy between $\Gamma$ and distance, 
particularly when $\Gamma$ is large and the distance estimate approaches the maximum limit, 
$d_{\rm max}$, at which $\beta \to 1$.  Only when $\Gamma$ is relatively modest and the
distance is well-constrained below $d_{\rm max}$ can the Lorenz factor be reliably bounded 
from above via kinematic data.

Although direct measurements provide only limited upper bounds for $\Gamma$, there 
are indirect arguments which suggest that Lorentz factors are modest, usually $\lesssim 5$, 
e.g., based on observations of radio-X-ray coupling \citep{Fender:06}
and of the deceleration of jets at late times as they expand 
\citep{Miller-Jones:06}.  Alternatively, it has been argued that microquasar jets may have
$\Gamma \gtrsim 10$,  based on observations of very narrow jet opening angles and the
assumption that the jet plasma expands freely at nearly the speed of light 
\citep{Miller-Jones:06,Miller-Jones:07}.  However, large values of $\Gamma$ 
may not be required to explain narrow jets if they are confined externally;
indeed, Miller-Jones et al. have provided evidence for such confinement in some systems.   

During one observation (see Section \ref{sect:jetmotion}), we observed a motion of 
$23.6\pm0.5$ \masd\ along a position angle of $130\deg$, consistent with the approaching 
jet in \GRS.  At the parallax distance of 8.6 kpc, this corresponds to 
$1.2 c$, where $c$ is the speed of light, making this source just marginally superluminal.   
The magnitude of the motion we measured agrees (exactly) with that measured
by \citet{Fender:99} for the approaching jet of $23.6\pm0.5$ mas d$^{-1}$, 
but our position angle differs significantly from their value of $142\deg\pm2\deg$, 
estimated from the orientation on the sky of multiple jet components separated by 
$\sim300$ mas in 1997 November.  Our measured motion also agrees with the average of 
those of \citet{Dhawan:00} of $22.2\pm1.3$ mas d$^{-1}$, but differs from their
average position angle of $141\deg\pm3\deg$  However, the magnitude of our measured motion  
is greater than the average of $17.3\pm0.2$ mas d$^{-1}$ toward $151\pm3\deg$ 
for the approaching jet observed from components separated by $\approx1000$ mas 
by \citet{Mirabel:94} and \citet{Rodriguez:99} in 1994/1995.  These jet parameters
are summarized in Table \ref{table:jetproperties}.

\begin{deluxetable}{lrrrl}
\tablecolumns{5} \tablewidth{0pc} 
\tablecaption{\GRS\ Jet Properties}
\tablehead {
 \colhead{Date} &\colhead{Angular Scale}&\colhead{P.A.}   &\colhead{$\mu_{app}$}  &\colhead{Reference} \\
 \colhead{}     &\colhead{(mas)}        &\colhead{(deg)}  &\colhead{(\masd)}      &\colhead{}
           }
\startdata
  1994/1995     &1000                   &$151\pm3$        &$17.3\pm0.2$      & \citet{Rodriguez:99} \\
  1997 November & 300                   &$142\pm2$        &$23.6\pm0.5$      & \citet{Fender:99} \\
  1997 October  &  50                   &$138\pm5$        &$22.0\pm2.0$      & \citet{Dhawan:00} \\
  1998 May      &  50                   &$143\pm4$        &$22.3\pm1.7$      & \citet{Dhawan:00} \\
  2013 May      &   5                   &$130\pm1$        &$23.6\pm0.5$      & this paper \\
\enddata
\tablecomments{\footnotesize
~Column 1 gives the approximate date of the observations; column 2 indicates the characteristic
angular scale of the observed jet; columns 3 and 4 give the position angle (east of north) and
the apparent proper motion of the approaching jet components.  Values from \citet{Rodriguez:99} 
are variance weighted means of 5 measurements listed in their Table 2.
The uncertainty in the P.A. of the jet for the \citet{Fender:99} observation is our estimate.
Values from \citet{Dhawan:00} are for observations where moving ejecta are seen. 
              }
\label{table:jetproperties}
\end{deluxetable}

Combining the measurements of the motion of the approaching and receding jet
components with the source distance allows one to calculate the inclination angle of the 
jet from the line of sight, $i$, and the bulk flow speed in the jet, $\beta$,
in units of the speed of light \citep{Fender:99}.  Using the parallax distance of 
$8.6^{+2.0}_{-1.6}$ kpc (Section \ref{sect:parallax}) and the apparent 
speeds of the approaching ($23.6\pm0.5$ \masy) and receding ($10.0\pm0.5$ \masy) 
jet components from Fender \etal, we find $i=60\pm5$ degrees and $\beta=0.81\pm0.04$.
This constrains the Lorentz factor $\gamma=(1-\beta^2)^{-1/2}$ to have values 
between 1.6 and 1.9.

There are now measurements of jet orientation and bulk flow speed over a wide
range of angular scales for \GRS\ (Table \ref{table:jetproperties}).  Clearly the
jet appears to bend through a position angle of 20\deg\ on the sky between scales of 
5 mas (40 AU) and 1\as0 (0.04 pc) from the black hole.  The apparent speed of the 
approaching jet was the same for our observations and that of \citet{Fender:99}.
However, \citet{Rodriguez:99} measured lower apparent speeds for both the approaching and 
receding jets compared to Fender \etal\ (see Table \ref{table:jetproperties}).  
While both jet orientation and internal flow speed can affect apparent motion, 
the ratio of approaching to receding motions measured by Rodr\'iguez \& Mirabel and 
Fender \etal\ are similar and most readily explained with a nearly constant inclination 
of $\approx60\deg$, but with different internal flow speeds of  $\beta\approx0.65$ and 
$\beta\approx0.81$, respectively.  Therefore, we conclude that either the jet bends and 
decelerates between radii of 0.01 to 0.04 pc, and/or plasma is ejected with different speeds 
at different times.

\subsection{Black Hole Mass} \label{sect:bhmass}

Our parallax distance of $8.6^{+2.0}_{-1.6}$ kpc is lower than the value of 11 kpc 
often adopted for \GRS.  However, it is consistent with the estimate of $\lax10$ kpc 
of \citet{Zdziarski:14}, which is based on a relation between jet power and luminosity.
Distance directly impacts estimates of the black hole mass \Mbh, based on orbital 
period and companion radial velocity measurements, which constrain $\Mbh \sin^3{i}$.  
Since the inclination of the system ($i$), as inferred from the apparent speeds of the
approaching ($\mu_a$) and receding ($\mu_r$) radio jets, is distance dependent 
(via ${\tan{i}\over d} = {2\over c} {{\mu_a\mu_r}\over{\mu_a-\mu_r}}$), 
a lower distance leads to a smaller binary inclination and a higher mass. 

\begin{figure}[ht]
\epsscale{0.8} 
\plotone{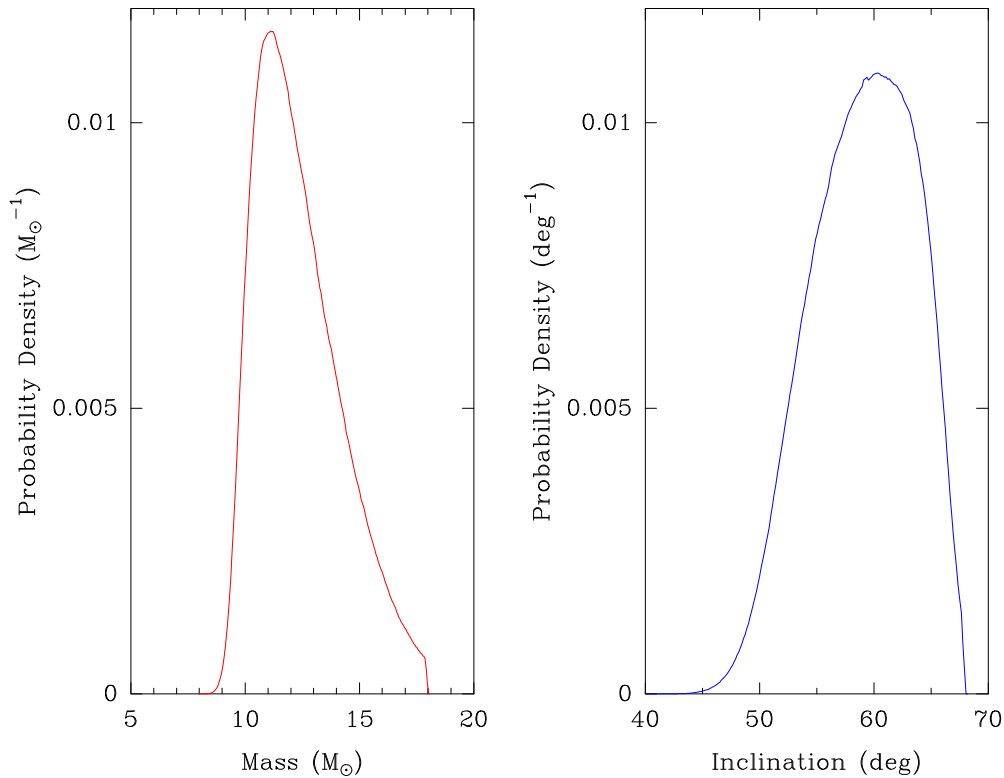}
\caption{\footnotesize
~Probability density function (PDF) for the \GRS\ black hole mass ({\it left panel}) 
and inclination of its spin axis ({\it right panel}), based on the parallax distance 
PDF shown in Fig. \ref{fig:distpdf}.  See Section \ref{sect:bhmass} for details.
         }
\label{fig:masspdf}
\end{figure}

\citet{Steeghs:13} assumed a uniform distance prior of $10 < d < 12$ kpc 
(and a direct relation between distance and inclination, which did not incorporate
uncertainty in the apparent jet motions), leading to an estimate of $\Mbh=10.1\pm0.6$ \Msun.  
In order to quantify the impact of our distance on the estimate of black hole mass, 
we revisited the Monte-Carlo simulations of Steeghs \etal\ and used our parallax distance 
PDF (see Fig.~\ref{fig:distpdf}) as prior information.  Each trial distance was combined with 
a trial for the observed jet velocity, based on the values and uncertainties of \citet{Fender:99}, 
to calculate the implied inclination.  This inclination was then combined with the spectroscopic 
parameters to calculate a black hole mass, contributing to the posteriori mass distribution.
The resulting mass and inclination distribution functions are shown in Fig.~\ref{fig:masspdf}.  
Both PDFs are significantly asymmetric, with extended tails towards higher mass and lower
inclination.  
The peak of the inclination PDF is at 60\deg\ with an expectation value of 59\deg\  
and a minimum-width 68\% confidence interval of 55\deg to 64\deg.
The peak of the mass PDF is at 11.2 \Msun\ with an expectation value of 12.4 \Msun\ 
and a minimum-width 68\% confidence interval of 10.5 to 14.3 \Msun, suggesting a slightly 
higher mass than quoted in Steeghs et al. 

Concerning X-ray luminosity, we note that our 8.6~kpc distance estimate
deflates claims for extreme values \citep[e.g., $\sim2-3~L_{\rm Eddington}$ 
for an assumed distance of $d=12.5$~kpc by][]{Done:04} and indicates that
the maximum luminosity of \GRS\ is at or near the Eddington limit.

\subsection{Black Hole Spin} \label{sect:bhspin}

\citet{McClintock:06} estimated the dimensionless spin parameter
\footnote{Spin is usually expressed in terms of the dimensionless spin parameter 
$a_* \equiv cJ/GM^2$, where $J$ and \Mbh\ are the angular momentum and mass of the 
black hole, respectively, and $|a_*| \le 1$},  
$a_*$, for the black hole in \GRS\ using the continuum-fitting method \citep{McClintock:13}
and a sample of seven selected spectra.  In applying this method,
one fits a thermal continuum spectrum to the thin disk model of \citet{Novikov:73}; 
the key parameter is the radius of the inner edge of the accretion disk,
assumed to be the innermost stable circular orbit for a black hole,  
which is directly related to \Mbh\ and $a_*$.  
The continuum-fitting method relies upon having independent and accurate estimates 
of distance, the black hole mass, and the inclination of its spin axis to the line of sight.
\citet{McClintock:06} assumed a distance of 11 kpc and adopted $\Mbh=14.0\pm4.4$ \Msun\ 
and $i=66\deg\pm2\deg$, which yielded an extreme value for the spin parameter of $a_* > 0.98$.  

Using our improved constraints on $d$, \Mbh and $i$, we present a preliminary analysis 
for the spin of \GRS\ (details will appear in a forthcoming paper).
We have analyzed a large sample of $\sim 2500$ {\it RXTE} spectra obtained with the 
PCU-2 detector, for which the quality of the data calibration has been significantly 
enhanced using the tool {\sc pcacorr} \citep{Garcia:14}.  We modeled the spectra using the 
relativistic disk model {\sc kerrbb2} \citep{McClintock:06,Li:05,Davis:06} and 
a cutoff Compton power-law component {\sc simplcut}, a variant of {\sc simpl} 
\citep{Steiner:09b}.  For a given set of values of $d$, \Mbh\ and $i$, a subset of 
the data is selected that meets standard criteria (e.g., Steiner et al. 2011): 
goodness of fit ($\chi^2_\nu<2$), low-to-moderate luminosity 
($L_{\rm disk} < 0.3 L_{\rm Eddington}$), 
and a dominant thermal component with scattering fraction $<25$\% \citep{Steiner:09a}.
The size of the data sample, selected over the broad grid of values of $d$, \Mbh\ and $i$, 
ranges from 8 to 167 spectra.  (The luminosity criterion, based on the values 
of $d$, \Mbh\ and $i$, determines the number of selected spectra.)  
For each set of parameters, a weighted mean value of spin is computed. 
Our preliminary estimate is $a_* = 0.98 \pm 0.01$, where the uncertainty is statistical only.
Our result is consistent with the original estimate of \citet{McClintock:06}, but 
it is now based firmly on a parallax measurement of distance and a much improved
estimate of the black hole's mass.
Including sources of systematic uncertainty \citep{Steiner:11}, we estimate 
$a_* > 0.92$ at $3\sigma$ confidence.

\subsection{Galactic Environment and Peculiar Motion}

\citet{Rodriguez:98} and \citet{Chaty:01} mapped a field of $1\deg$ centered near 
\GRS\ at cm-wavelengths and found 
two \HII\ regions (\IRASS\ and \IRASN) symmetrically offset by $17'$ about \GRS\ along a
position angle of about $157\deg\pm1\deg$ (our measurement from their published figures).  
They also found a nonthermal ``trail'' of emission near \IRASS\ along a position angle
of about $152\deg\pm5\deg$, which points back toward \GRS.  However, since the \HII\ 
regions have (far) kinematic distances between about 6 and 8 kpc, and they assumed a distance 
of 12.5 kpc for \GRS, Rodr\'iguez \& Mirabel reluctantly concluded that these sources were 
only a chance projection.  However, our near-zero parallax of \GRS\ relative to \Mref,
and the likely association of the \HII\ regions (\IRASS\ and \IRASN) and the water masers
(\Mref\ and \Mcheck) with massive star formation in the Sagittarius spiral arm of the
Galaxy \citep{Wu:14}, now suggests that most or all of these sources are spatially 
related.  In this case, the nonthermal trail of emission from \IRASS\ pointing toward \GRS\
could indeed be related to its jet emission, as modeled by \citet{Kaiser:04}.

The peculiar velocity components for \GRS\ estimated in Section \ref{sect:propermotion}
are relatively modest and consistent with an old stellar system that has orbited 
the Galaxy many times, having responded to random gravitational perturbations from 
encounters with spiral arms and giant molecular clouds, and acquired a peculiar velocity
of 20-30 \kms\ as observed.  This suggests that the \GRS\ binary system did not receive a 
large velocity ``kick,'' owing to mass loss from the primary as it evolved and became 
a black hole.

\vskip 0.5truein 
JFS was supported by NASA Hubble Fellowship grant HST-HF-51315.01, and the
work of JEM was supported in part by NASA grant NNX11AD08G.

\vskip 0.5truein 
{\it Facilities:}  \facility{VLBA}


\begin{thebibliography}{}
\bibitem[Chapuis \& Corbel(2004)]{Chapuis:04}
         Chapuis, C. \& Corbel, S. 2004, \aap, 414, 659
\bibitem[Chaty \etal(2001)]{Chaty:01}
         Chaty, S., Rodr\'guez, L. F., Mirabel, I. F. \etal 2001, \aap, 366, 1035
\bibitem[Davis \& Hubeny(2006)]{Davis:06}
         Davis, S.~W. \& Hubeny, I. 2006, \apjs, 164, 530
\bibitem[Dhawan, Mirabel \& Rodr\'iquez(2000)]{Dhawan:00}
         Dhawan, V., Mirabel, I. F. \& Rodr\'iguez, L. F. 2000, \apj, 543, 373
\bibitem[Dhawan \etal(2007)]{Dhawan:07}
         Dhawan, V., Mirabel, I. F., Rib\'o, M. \& Rodrigues, I. 2007, \apj, 668, 430
\bibitem[Done, Wardzi\'nski \& Gierli\'nski(2004)]{Done:04}
         Done, C., Wardzi\'nski, G. \& Gierli\'nski, M. 2004, \mnras, 349, 393
\bibitem[Garcia \etal(2014)]{Garcia:14}
         Garcia, J., McClintock, J.~E., Steiner, J.~F., Remillard, R.~A., 
         \& Grinberg, V.\ 2014, to appear in \apj\ (arXiv:1408:3607 [astro-ph.HE]) 
\bibitem[Greiner \etal(2001)]{Greiner:01}
         Greiner, J., Cuby, J. G., McCaughrean, M. J., Castro-Tirado, A. J. \&
         Mennickent, R. E. 2001, \aap, 373, L37
\bibitem[Fender \etal(1999)]{Fender:99}
         Fender, R. P., Garrington, S. T., McKay, D. J. \etal\ 1999, \mnras,304, 865
\bibitem[Fender(2003)]{Fender:03}
         Fender, R.~P. 2003, \mnras, 340, 1353
\bibitem[Fender \etal(2004)]{Fender:04}
         Fender, R.~P., Belloni, T.~M., \& Gallo, E. 2004, \mnras, 355, 1105
\bibitem[Fender(2006)]{Fender:06}
         Fender, R.~P. 2006, in ``Compact stellar X-ray sources'', 
         Edited by Walter Lewin \& Michiel van der Klis. Cambridge Astrophysics Series, No. 39. 
         Cambridge, UK: Cambridge University Press, p381
\bibitem[Kaiser \etal(2004)]{Kaiser:04}
         Kaiser, C. R., Gunn, K. f., Brocksopp, C. \& Sokoloski, J. L. 2004, \apj, 612, 332
\bibitem[Li \etal(2005)]{Li:05}
         Li, L.-X., Zimmerman, E.~R., Narayan, R. \& McClintock, J.~E. 2005,
         \apjs, 157, 335
\bibitem[McClintock \etal(2006)]{McClintock:06}
         McClintock, J.~E., Shafee, R., Narayan, R., Remillard, R.~A., Davis, S.~W., 
         \& Li, L.-X. 2006, \apj, 652, 518
\bibitem[McClintock \etal(2013)]{McClintock:13}
         McClintock, J.~E.,  Narayan, R., \& Steiner, J.~F.\ 2013, \ssr, 73
\bibitem[Miller-Jones, Fender \& Nakar(2006)]{Miller-Jones:06}
         Miller-Jones, J.~C.~A., Fender, R.~P., \& Nakar, E.\ 2006, \mnras, 367, 1432 
\bibitem[Miller-Jones \etal(2007)]{Miller-Jones:07}
         Miller-Jones, J.~C.~A., Rupen, M.~P., Fender, R.~P., et al.\ 2007, \mnras, 375, 1087
\bibitem[Miller-Jones \etal(2009)]{Miller-Jones:09}
         Miller-Jones, J.~C.~A., Jonker, P. G., Dhawan, V., et al.\ 2009, \apjl, 706, L230
\bibitem[Mirabel \& Rodr\'iguez(1994)]{Mirabel:94}
         Mirabel, I. F. \& Rodr\'iguez, L. F. 1994, \nat, 371, 46
\bibitem[Novikov \& Thorne(1973)]{Novikov:73}
         Novikov, I.~D., \& Thorne, K.~S.\ 1973, Black Holes (Les Astres Occlus), 343 
\bibitem[Reid \etal(2009)]{Reid:09}
         Reid, M. J., Menten, K. M., Brunthaler, A., Zheng, X. W., Moscadelli, L. \&
         Xu, Y. 2009, \apj, 693, 397
\bibitem[Reid \etal(2011)]{Reid:11}
         Reid, M. J., McClintock, J. E., Narayan, R., Gou, L., Remillard, R. A. \&
         Orosz, J. A. 2011, \apj, 742, 83
\bibitem[Reid \& Honma(2014)]{Reid:13}
         Reid, M. J. \& Honma, M. 2014, to appear in \araa\ (arXiv1312.2871)
\bibitem[Reid \etal(2014)]{Reid:14}
         Reid, M. J., Menten, K. M., Brunthaler, A., \etal\ 2014, \apj, 783, 130
\bibitem[Rodr\'iguez \& Mirabel(1998)]{Rodriguez:98}
         Rodr\'iguez, L. F. \& Mirabel, I. F. 1998, \aap, 340, L47
\bibitem[Rodr\'iguez \& Mirabel(1999)]{Rodriguez:99}
         Rodr\'iguez, L. F. \& Mirabel, I. F. 1999, \apj, 511, 398
\bibitem[Steeghs \etal(2013)]{Steeghs:13}
         Steeghs, D., McClintock, J. E., Parson, S. G., Reid, M. J., Littlefair, S. \&
         Dhillon, V. S. 2013, \apj, 768, 185
\bibitem[Steiner \etal(2009a)]{Steiner:09a}
         Steiner, J.~F., McClintock, J.~E., Remillard, R.~A., Narayan, R., \&
         Gou , L.~J. 2009{\natexlab{a}}, \apjl, 701, L83
\bibitem[Steiner \etal(2009b)]{Steiner:09b}
         Steiner, J.~F., Narayan, R., McClintock, J.~E., \& Ebisawa, K.
         2009{\natexlab{b}}, \pasp, 121, 1279
\bibitem[Steiner \etal(2011)]{Steiner:11}
         Steiner, J.~F., Reis, R.~C., McClintock, J.~E., Narayan, R.,
         Remillard , R.~A.,  Orosz , J.~A.,  Gou, L., Fabian, A.~C., \& Torres, M.~A.~P. 
         2011, \mnras, 416, 941
\bibitem[Steiner \& McClintock(2012a)]{Steiner:12a}
         Steiner, J.~F., \& McClintock, J.~E. 2012, \apj, 745, 136 
\bibitem[Steiner, McClintock \& Reid(2012b)]{Steiner:12b}
         Steiner, J.~F., McClintock, J.~E. \& Reid, M. J. 2012, \apjl, 745, L7 
\bibitem[Wu \etal(2014)]{Wu:14}
         Wu, Y., Sato, M., Reid, M. J., \etal 2014, \aap, 566, 17
\bibitem[Zdziarski (2014)]{Zdziarski:14}
         Zdziarski, A. A. 2014, \mnras, 444, 1113 
    
\end{thebibliography}
\end{document}